\documentclass{article}
\usepackage{spconf,amsmath,graphicx,amsfonts}

\usepackage{enumitem}
\setlist{nosep, leftmargin=14pt}

\usepackage{mwe} % to get dummy images

\title{ToothSegNet: Image Degradation meets Tooth Segmentation in CBCT Images}

\name{Jiaxiang Liu$^1$, Tianxiang Hu$^1$, Yang Feng$^2$, Wanghui Ding$^3$, Zuozhu Liu$^{1, 3 *}$ \thanks{This work is supported by the National Natural Science Foundation of China (Grant No. 62106222), the Natural Science Foundation of Zhejiang Province, China(Grant No. LZ23F020008) and the Zhejiang University-Angelalign Inc. R$\&$D Center for Intelligent Healthcare.
}}
\address{1) Zhejiang University-University of Illinois at Urbana-Champaign Institute, Zhejiang University\\
	2) Angelalign Inc.\\
        3) Stomatology Hospital, School of Stomatology, Zhejiang University School of Medicine  }

\begin{document}
%\ninept
%
\maketitle
\begin{abstract}
In computer-assisted orthodontics, three-dimensional tooth models are required for many medical treatments. Tooth segmentation from cone-beam computed tomography (CBCT) images is a crucial step in constructing the models. However, CBCT image quality problems such as metal artifacts and blurring caused by shooting equipment and patients’ dental conditions make the segmentation difficult. In this paper, we propose ToothSegNet, a new framework which acquaints the segmentation model with generated degraded images during training. ToothSegNet merges the information of high and low quality images from the designed degradation simulation module using channel-wise cross fusion to reduce the semantic gap between encoder and decoder, and also refines the shape of tooth prediction through a structural constraint loss. Experimental results suggest that ToothSegNet produces more precise segmentation and outperforms the state-of-the-art medical image segmentation methods.
% To deal with the low-quality input in practice, we propose a tooth segmentation method termed ToothSegNet which improves the robustness of the segmentation model by acquainting the model with generated degraded image samples. During training, ToothSegNet merges the information of high and low quality images from the degradation simulation module using a channel-wise cross fusion module to reduce the semantic gap between the encoder and decoder, and also updates according to a structural constraint loss that refines the shape of teeth. 
\end{abstract}
\section{Introduction}
\label{sec:intro}
Stomatologists and dentists use tooth models to carry out diagnosis, orthodontic treatment planning and dental restoration \cite{cui2022fully}. Constructing a tooth model first requires tooth segmentation from CBCT scans \cite{weiss2019cone}. Existing computational methods then generate the final tooth models according to the segmentation. Traditionally, specialists have to manually label each tooth from CBCT images slice by slice, which is a huge workload and extremely time-consuming. It is therefore practically demanded to design accurate and fully automatic end-to-end methods to attain tooth segmentation from CBCT images \cite{hao2022ai}. 

In recent years, deep learning has been increasingly applied to image segmentation tasks in various medical areas including dentistry \cite{weiss2019cone}. Some prior works \cite{cui2019toothnet}, \cite{wu2020center} formulated CBCT tooth segmentation task as 3D instance segmentation which usually demands labeling tooth pixels and instances on 3D voxels across the entire CBCT scan. To explore a more efficient way, we recognize the task as semantic segmentation over 2D CBCT images to alleviate the data annotation labour. Directly applying current segmentation methods to the 2D case \cite{ronneberger2015u}, however, leads to unsatisfied results because of the existence of low-quality images.
% For instance, if a patient has a metal insert such as dental fillings, orthopaedic implants or other metal implants in the mandible, since the attenuation coefficient of the internal tissue of the human body is much lower than that of the embedded metal object, the rays will be severely attenuated after penetrating the metal embedded object, resulting in metal artifacts in CBCT images. Another common defect is image blurring which is usually caused by patient movement during CT scanning. 
Fig.1 illustrates two most commonly seen defects in dental CBCT images  \cite{cui2021hierarchical}: (a) consecutively displays a high quality CBCT image, an image with metal artifacts and an image with blurring; (b) and (c) give respective ground truth and segmentation results of U-Net \cite{ronneberger2015u}; the red boxes are magnification areas for metal artifacts and blurring. The second and third triplets indicate that such image defects can lead to abysmal tooth segmentation. Addressing these problems should greatly improve 2D CBCT tooth segmentation, promising an efficient and effective tooth modeling for the dentistry community. 
% As shown in the second triplet, with metal artifacts, it is very likely for the present segmentation methods to misclassify pixels, resulting in tooth adhesion and vague boundaries in tooth prediction. The third triplet indicates that image blurring can lead to abysmal segmentation performance on tiny teeth. Addressing these problems should greatly improve automatic tooth segmentation and promise a more effective tooth modeling for the dentistry community. 
\begin{figure*}
\centering
% Metal artifacts and blurring in CBCT images. The first row are the CBCT source images; The second row are the ground truth of CBCT images; The third row are the segmentation results of CBCT images by Unet method.
\includegraphics[width=1\textwidth]{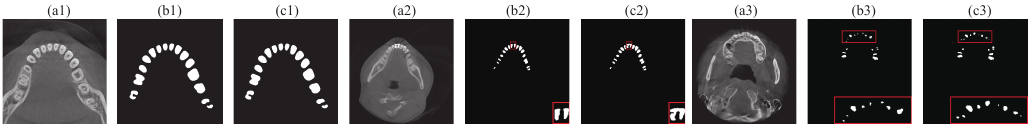}
\caption{Metal artifacts and blurring in CBCT images.} \label{fig1}
\setlength{\abovecaptionskip}{-10cm}
\vspace{-15pt} %调整图片与上文的垂直距离
\end{figure*}

In this study, we propose a metal artifacts and blurring robust and teeth structural constrained tooth segmentation method named ToothSegNet. The philosophy is to acquaint the deep model with defective cases when learning each CBCT image sample. Experiments show that our method not only outperforms the state-of-the-art segmentation methods on high-quality images but also has a strong robustness to images with metal artifacts and blurring, suggesting the applicability of our framework in real-world clinical scenarios. The main contributions are summarized below:
\begin{itemize}
    \item We propose a medical image degradation simulation strategy, combined with a multi quality fusion module, to attain higher robustness to problematic images. 
    \item We design a channel-wise cross fusion module (CCF) to reduce the semantic gap between the encoder and decoder, which eliminates the ambiguity features existed in vanilla U-Net caused by skip-connection.
    \item We design a structural constraint loss in order to restrict the structure of tooth prediction.
    \item We explore an annotation approach which is much lower-cost than that implemented in current CBCT tooth segmentation methods.
    % integrate it with other designs to propose ToothSegNet.
\end{itemize}

\section{Method}
\label{sec:method}

\subsection{Overview}
% Let's define the tooth segmentation task for CBCT images concretely.
Suppose we have a dataset $\mathcal{D} = \{(\mathbf{x}_i, \mathbf{y}_i)\}_{i=1}^N$, where $\mathbf{x}_i$ and $\mathbf{y}_i$ denote the $i$-th CBCT image and corresponding  pixel-level segmentation labels respectively, $N$ is the number of images in the dataset. For each pixel, we use $y_i \in \{0, 1\}$ to denote the background and tooth respectively. In our task, we aim to generate accurate pixel-level predictions  $\mathbf{\hat{y}}_i$ for each input image. As is illustrated in Fig. 2,  ToothSegNet is composed of the degradation simulation module, multi-quality fusion module, CCF, vanila U-Net decoder, SE-Net \cite{hu2018squeeze} and the structural constraint loss. The vanila U-Net decoder follow the conventional CNN-based architecture \cite{ronneberger2015u}.
% During training, each CBCT image $\mathbf{x}_i$ is performed a degradation simulation to randomly generate a degraded image $\mathbf{x}_i^d$. $\mathbf{x}_i$ and $\mathbf{x}_i^d$ will be passed through the multi-quality fusion module together, and the fused features are then sent to SE-Net [16] for channel weight allocation as well as CCF for source image and degraded image feature fusion. Finally, CCF encoded features and SE-Net enhanced features fuse into one, conveyed to vanilla U-Net decoder to get the output. We also designed a structural constraint loss function to refine the shape of prediction to attain clearer tooth boundaries.  

\subsection{Degradation simulation module}
% As illustrated in Fig. 1, current deep learning based segmentation methods do not perform well on both metal artifact and blurring cases. One natural analysis is that these models are not fully exposed to degraded CBCT images during training. In this regard, therefore, we designed a module to generate low-quality degraded image samples, and a training pipeline to familiarize the model with these images.
% In computer vision tasks on natural images, image degradation often is used to construct super-resolution, defogging or deblurring-based datasets in order to simulate complex situations in natural scenes\cite{yang2019deep}, \cite{tian2020coarse}, \cite{pei2019effects}.  The designed ToothSegNet introduces the idea to degrade the quality of the medical source images for blurring and metal artifact simulation.
In computer vision tasks, people often construct data sets through image degradation to mimic complex real-world scenarios \cite{tian2020coarse}, \cite{pei2019effects}. Referred to that, ToothSegNet randomly degrades the input image during training. For blurring and double blurring simulation, the input image is down-sampled once (1x) and twice (2x) respectively and then up-sampled to generate the degraded image $\mathbf{x}_i^d$, as is shown in Fig. 2. In the downsampling, a Gaussian pyramid blurring is performed to downscale the image. In the upsampling, a bilinear interpolation is performed to convert images to the original size. For artifact simulation, since we observe that the teeth with metal have higher image contrast compared with others, we create contrast enhanced images $\mathbf{x}_i^d$ by image square. 

The degradation simulation procedure is defined to follow the operation probability distribution:
% The original code reporting "undefined control sequence" error.
% \begin{equation}
% \setlength{\abovedisplayskip}{6pt}
% {\text{\mathbb{P}}}\left( {X_i^d = {\phi ^{(j)}}(\mathbf{x}_i)\mid {X_i} = \mathbf{x}_i} \right) = \frac{1}{4} \ \ \forall j \in \{ 1,2,3,4\} 
% \setlength{\belowdisplayskip}{6pt}
% \end{equation}
% \begin{equation}
% \setlength{\abovedisplayskip}{6pt}
% \text{{\mathbb{P}}}\left( {X_i^d = {\phi ^{(j)}}(\mathbf{x}_i)\mid {X_i} = \mathbf{x}_i} \right) = \frac{1}{4} \ \ \forall j \in \{ 1,2,3,4\} 
% \setlength{\belowdisplayskip}{6pt}
% \end{equation}
\begin{equation}
\setlength{\abovedisplayskip}{6pt}
{\mathbb{P}}\left( {X_i^d = {\phi ^{(j)}}(\mathbf{x}_i)\mid {X_i} = \mathbf{x}_i} \right) = \frac{1}{4} \ \ \forall j \in \{ 1,2,3,4\} 
\setlength{\belowdisplayskip}{6pt}
\end{equation}
where $X_{i}$ and $X_i^d$ denote an input CBCT image and its degradation simulation, ${\phi ^{(1)}}$ to ${\phi ^{(4)}}$ represent the blurring degradation, double blurring degradation, no operation for source images and artifact degradation respectively.

\begin{figure*}
\centering

\includegraphics[width=\textwidth]{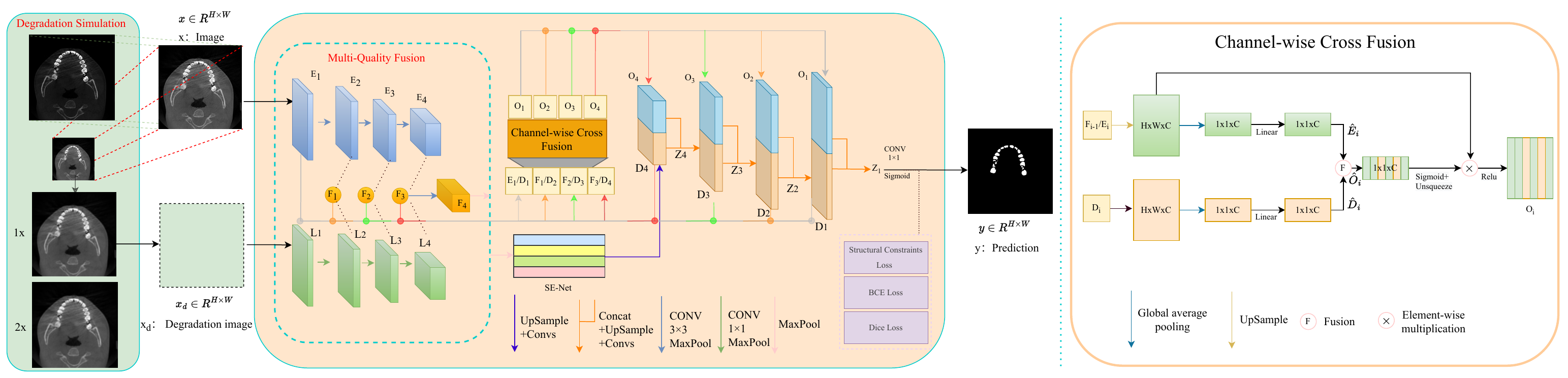}
\caption{The pipeline of ToothSegNet and details of the channel-wise cross fusion module.} \label{fig1}
\vspace{-15pt} %调整图片与上文的垂直距离
\end{figure*}

\subsection{Network architecture}
% As is illustrated in Fig.2, ToothSegNet is composed of the degradation simulation module, multi-quality fusion module, CCF, vanila U-Net decoder, and SE-Net \cite{hu2018squeeze}. The vanila U-Net decoder follow the conventional CNN-based architecture \cite{ronneberger2015u}. 
% %The feature map sizes of encoder and decoder are shown in Table 1.
\subsubsection{Multi-quality fusion module}
After the degraded images are obtained, the source and degraded images are sent to a designed multi-quality fusion module similar to the two-branch improved U-Net encoder. As shown in Fig. 2, the blue branch is the vanilla U-Net encoder which encodes the feature of source images, the green branch is the U-Net encoder with 1x1 convolution which encodes the feature of degraded images. The fused feature is defined as follows:
\begin{equation}
\setlength{\abovedisplayskip}{6pt}
{F_i} = {E_{i + 1}} + {L_{i + 1}} \ \ \ {\text{      i}} \in \{ 1,2,3\} 
\setlength{\belowdisplayskip}{6pt}
\end{equation}
$F_4$ is the fused feature.

\subsubsection{Channel-wise cross fusion}
Shallow layer features with less semantic information may damage performance via a skip connection in conventional U-Net \cite{wang2021uctransnet}. To solve this issue, we design the CCF module, similar to channel attention \cite{hu2018squeeze}, \cite{wang2021uctransnet}, to increase the semantic gap between teeth and the background. Fig. 2 shows the framework of CCF. The encoded CCF feature is defined as:
\begin{equation}
\setlength{\abovedisplayskip}{6pt}
{O_i} = \left\{ {\begin{array}{*{20}{l}}
  {{\text{CCF}}\left( {{E_1},{D_1}} \right)}&{i = 1} \\ 
  {\operatorname{CCF} \left( {{F_{i - 1}},{D_i}} \right)}&{i = 2,3,4} 
\end{array}} \right.
\setlength{\belowdisplayskip}{6pt}
\end{equation}
The feature fusion of CCF in Fig.2 is defined as:
\begin{equation}
\setlength{\abovedisplayskip}{6pt}
    \hat{O}_i = \hat{E}_i + \hat{D}_i
\setlength{\belowdisplayskip}{6pt}
\end{equation}
%x = α x_1 + β x_2其中埃尔法和贝塔上可学习参数。其中， α和β是可学习参数。
% 这段还没改完 Need some discussion
Also, the SE-Net module is added to exploit the correlation of feature channels, enhancing the feature representation:
\begin{equation}
\setlength{\abovedisplayskip}{6pt}
{D_i} = \left\{ {\begin{array}{*{20}{l}}
  {{\text{concat}(D_{i+1},O_{i+1})}}&{i = 1,2,3} \\ 
  {{\mathcal{F}_{{\text{SE}}}}{\text{(}}{{{F}}_5})}&{i = 4} 
\end{array}} \right.,
\setlength{\belowdisplayskip}{6pt}
\end{equation}
where ${{\mathcal{F}}_{\text{SE}}}$ denotes the SE-Net. ${{\mathcal{D}}_{\text{i}}}$ are the fused features of inconsistent semantics between the CCF and SE-Net.

% \begin{table}
% \centering
% \caption{The sizes of encoder and decoder}
% \label{fig4}
% \resizebox{.5\textwidth}{!}{
% \begin{tabular}{l|l|l|l|l} 
% \hline
% $E_1 \in R^{64* {H}{}*{W}{}}$ & $L_1 \in R^{64* {H}{}*{W}{}}$ & $F_1 \in R^{128* \frac{H}{2}*\frac{W}{2}}$ & $D_1 \in R^{64* \frac{H}{2}*\frac{W}{2}}$ & $Z_1 \in R^{64* {H}{}*{W}{}}$  \\
% $E_2 \in R^{128* \frac{H}{2}*\frac{W}{2}}$ & $L_2 \in R^{128* \frac{H}{2}*\frac{W}{2}}$ & $F_2 \in R^{128* \frac{H}{2}*\frac{W}{2}}$  & $D_2 \in R^{128* \frac{H}{4}*\frac{W}{4}}$ & $Z_2 \in R^{64* \frac{H}{2}*\frac{W}{2}}$  \\
% $E_3 \in R^{256* \frac{H}{4}*\frac{W}{4}}$ &$L_3 \in R^{256* \frac{H}{4}*\frac{W}{4}}$ & $F_3 \in R^{128* \frac{H}{2}*\frac{W}{2}}$  & $D_3 \in R^{256* \frac{H}{8}*\frac{W}{8}}$ & $Z_3 \in R^{128* \frac{H}{4}*\frac{W}{4}}$  \\
% $E_4 \in R^{512* \frac{H}{8}*\frac{W}{8}}$ &$L_4 \in R^{512* \frac{H}{8}*\frac{W}{8}}$ & $F_4 \in R^{128* \frac{H}{2}*\frac{W}{2}}$  & $D_4 \in R^{512* \frac{H}{16}*\frac{W}{16}}$ & $Z_4 \in R^{256* \frac{H}{8}*\frac{W}{8}}$ \\
% \hline
% \end{tabular}
% }
% \end{table}

\subsection{Structural constraint loss}
With noise in images, the inference often lose the tooth structure. Thus, we design the structural constraint loss based on the structural similarity index measure (SSIM) \cite{wang2004image}: 
% SSIM is defined as follows:
% \begin{equation}
% \operatorname{SSIM}(x_1,x_2)=\frac{\left( 2{{\mu }_{x_1}}{{\mu }_{x_2}}+{{c}_{1}} \right)\left( 2{{\sigma }_{x_1x_2}}+{{c}_{2}} \right)}{\left( \mu _{x_1}^{2}+\mu _{x_2}^{2}+{{c}_{1}} \right)\left( \sigma _{x_1}^{2}+\sigma _{x_2}^{2}+{{c}_{2}} \right)},
% \end{equation}
% where ${\mu }_{x_1}$ and ${\mu }_{x_2}$ are mean values of $x_1$ and $x_2$, $\sigma _{x_1}^{2}$ and $\sigma _{x_2}^{2}$ are the variances of $x_1$ and $x_2$, ${{\sigma }_{x_1x_2}}$ is the covariance of $x_1$ and $x_2$, $c_1$ and $c_2$ are two variables to stabilize the division with weak denominator. 
\begin{equation}
\setlength{\abovedisplayskip}{6pt}
Los{s_{str}}{\rm{ = 1 -}} \frac{{SSIM\left( {I_{O}^{},\;I_{GT}^{}} \right) + SSIM\left( {I_C^{},\;I_{GT}^{}} \right)}}{2},
\setlength{\belowdisplayskip}{6pt}
\end{equation}
where $I_{GT}$ is the ground truth zero-one vector, $I_{O}$ is the prediction of ToothSegNet, $I_C^{}$ is the element-wise multiplication between $I_{O}$ and $I_{GT}$ which, as shown in Fig. 3, retains the correctly predicted area in $I_{O}$ and sets the wrong predicted area to 0. $Los{s_{str}}$ ensures that the prediction approaches $I_{GT}$ structurally, effectively constraining the results.

The total loss function is the sum of three losses:
\begin{equation}
\setlength{\abovedisplayskip}{3pt}
Los{s_{total}}{\rm{ = }} L{\rm{oss}}_{Dice}^{} +  Los{s_{BCE}} +  Los{s_{str}},
\setlength{\belowdisplayskip}{3pt}
\end{equation}
Besides $Los{s_{str}}$, $L{\rm{oss}}_{Dice}^{}$ is based on the Dice metrics \cite{minaee2021image} and $Los{s_{BCE}}$ is the binary cross-entropy (BCE) loss \cite{minaee2021image}.  
\begin{figure}
\includegraphics[width=0.5\textwidth]{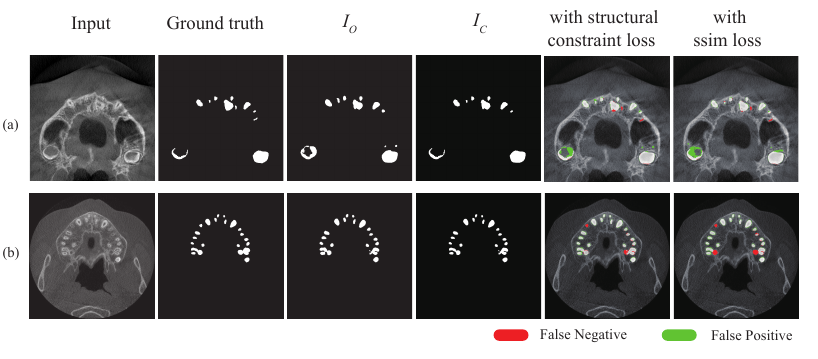}
\caption{The visualization of structural constraint loss.} \label{fig4}
\vspace{-20pt} %调整图片与上文的垂直距离
\end{figure}
\subsection{Training details}
We set the batch size to 4 and the epoch to 300. The input resolution and patch size are set as 512 x 512 and 16 respectively. For optimizer, we use Adam with the initial learning rate to be 0.01. We perform a number of data augmentations including horizontal flipping, vertical flipping and random rotating. 
% For results demonstration, we set the threshold to 0.8 (i.e., a pixel with the prediction larger than 0.8 in the final prediction map will be considered as a tooth pixel). 

\section{Experiment}
\label{sec:experiment}

\subsection{Dataset}
We construct a large-scale CBCT dataset consisting of 503 patient samples from hospitals and clinics across 25 provinces in China during 2018-2021. The 503 patients are at the age of 19.53±7.57 years old, with 32.5\% male and 67.5\% female. Each patient has 400 slices, from which we select 15-25 slices to annotate to make sure labeled images contain different anatomical information. Selected images are annotated with software Labelme, under the supervision of senior radiologists with more than 10 year experience. Accordingly, the dataset has 9651 CBCT images in total from which 8612 images of 453 patients are used to train the model and 1039 images of 50 patients are used to test the model.

\begin{table}
\centering
\caption{Patient level segmentation performance of ToothSegNet and baselines.}
\resizebox{.5\textwidth}{!}{
\begin{tabular}{c|cccc} 
\hline
            & \ \  IoU &  Dice  & Recall & Precision  \\ 
\hline
U-Net\cite{ronneberger2015u}       &  \underline{81.12±5.18} & \underline{88.95±3.54} &  94.63±4.12 & 84.82±5.35      \\
U-Net++\cite{zhou2018unet++}     &  80.70±5.53 & 88.61±3.78 & 93.17±4.40  & 85.46±5.40      \\
FCN \cite{long2015fully}        &  78.82±6.98 &  87.32±4.95 &  90.56±5.73 & 85.42±5.84      \\
DeepLabv3\cite{chen2017rethinking}   &  78.79±6.90 & 87.29±4.88 & 90.47±5.67  & 85.49±5.88      \\
MeDT\cite{valanarasu2021medical}        &  73.87±13.8  & 82.35±12.6& 76.98±14.3  & \textbf{94.42±4.91}      \\
UCTransNet\cite{wang2021uctransnet}  &  76.86±7.09 & 85.37±4.81 & \textbf{97.27±2.90}  & 78.70±7.83      \\ 
\hline\hline
ToothSegNet & \textbf{82.24±6.38}  & \textbf{89.74±4.07} & \underline{95.71±3.83}  & \underline{85.51±7.66}      \\
\hline
\end{tabular}
}
\vspace{-15pt} %调整图片与上文的垂直距离
\end{table}
\begin{figure*}
\centering
\includegraphics[width=0.8\textwidth]{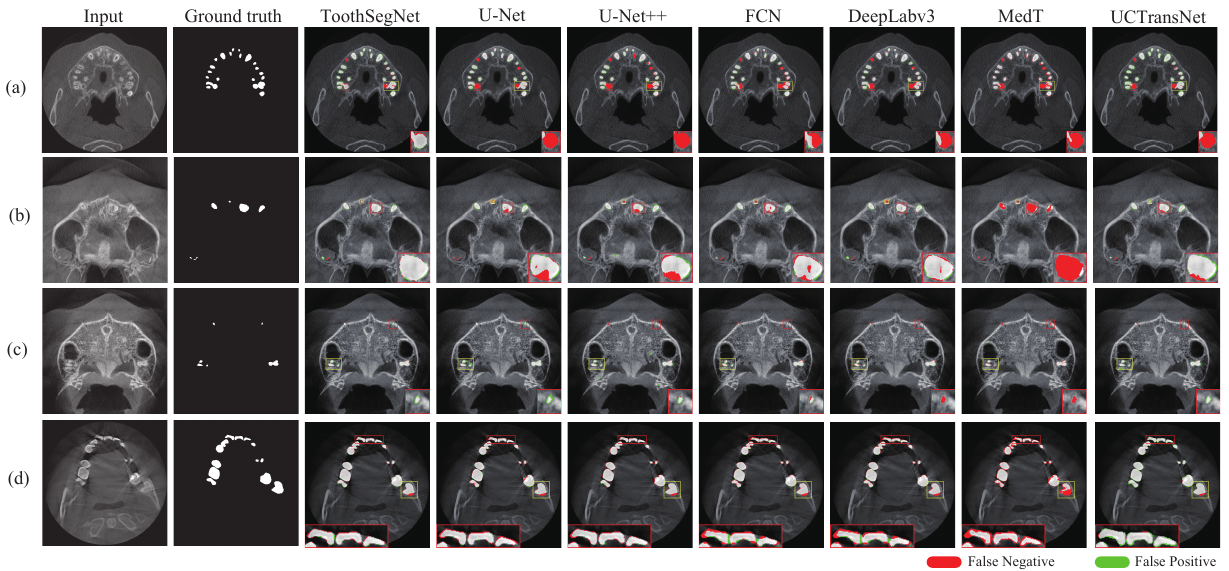}
\caption{Visualization of the segmentation result of different methods.} \label{fig3}
\vspace{-15pt} %调整图片与上文的垂直距离
\end{figure*}
\subsection{Main results}
To demonstrate the overall segmentation performance of the proposed ToothSegNet, we compare it with six methods for a comprehensive evaluation, covering four U-Net based methods: U-Net \cite{ronneberger2015u}, UNet++ \cite{zhou2018unet++}, FCN \cite{long2015fully}, Deeplabv3 \cite{chen2017rethinking} and two state-of-the-art transformer-based segmentation methods: MedT \cite{valanarasu2021medical}, and UCTransNet \cite{wang2021uctransnet}. To make a fair comparison, the implementations of MedT, UCTransNet, UNet, and UNet++ are based on their original source codes, and the implementations of FCN, and Deeplabv3 are based on the codes from MMsegmentation \cite{xu2020mmsegmenation} where their originally published settings are used in the experiment. The backbone of DeepLabv3 and FCN is ResNet-50, and the training hyper parameters are in MMSegmentation \cite{xu2020mmsegmenation} by default, the training procedure lasts for 160K iterations. Experimental results are reported in Table 1 where the best results are boldfaced, and the second results are underlined.

Each entry of Table 1 exhibits the average over 50 patient samples. The results indicate that ToothSegNet attains noticeable improvements over prior arts. In clinics, the segmentation results of tooth CBCT images need to be reconstructed into 3D mesh images, which turns out to require a balance between false positives and false negatives. Therefore, CBCT segmentation favors higher Dice and IoU. ToothSegNet achieves 89.74\% Dice and 82.24\% IoU with testing, which surpasses UCTransNet by 4.37\% Dice, and 5.38\% IoU. 

We visualize the segmentation results of all models in Fig. 4 where The red boxes contains FN pixels(red) and FP pixels(green). Fig. 4 (a) shows the results of a high-quality CBCT image on which ToothSegNet outperforms other methods. (b), (c) show the results of CBCT images with blurring, illustrating that our method generates segmentation results more similar to the ground truth than others. (d) shows the results of a CBCT image with metal artifacts, demonstrating that ToothSegNet have the best performance on tooth boundaries. 
% In summary, the proposed method preserves more information of teeth and reduces the confusing false-positive lesions which leads to clearer predictions around the boundaries.
% To test ToothSegNet's performance on intractable cases, low-quality images containing blurring or metal artifacts from 30 intractable samples are selected form test set. As is shown in Table III, ToothSegNet achieves 95.71\% Dice and 91.77\% IoU, which surpasses the second method testing results by 1.56\% Dice, and 2.82\% IoU.
%为了验证本文在困难数据中的表现，从外部测试集中选取30个模糊或带有金属伪影的低质量成像的病例进行验证。验证结果表明本文在困难样本上的表现更加优秀。
\begin{figure}
\centering
\includegraphics[width=0.5\textwidth]{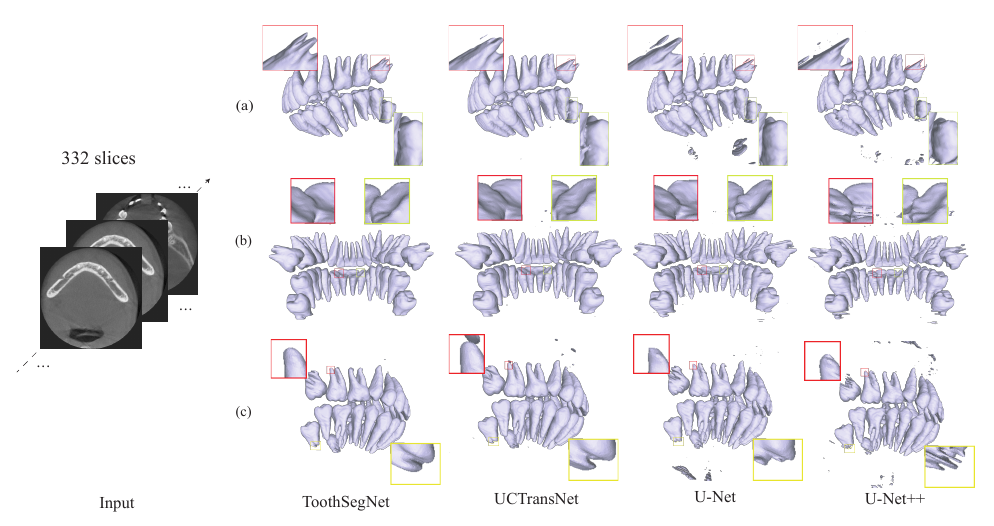}
\caption{The Visualization of reconstruction results. (a), (b) and (c) denotes right view, frontal view and left view.} \label{abl}
\vspace{-20pt} %调整图片与上文的垂直距离
\end{figure}
\subsection{Reconstruction}
To explore the performance of ToothSegNet in clinics, we employ the marching cubes algorithm \cite{hao2022ai} to get 3D CBCT mesh. Fig. 5 shows the reconstruction results using four state-of-the-art methods for a selected sample in which CBCT images are blurred around teeth erupting areas, and have high contrast differences as well as significant metal artifacts in teeth crown areas. As is shown in the right view, the tooth root reconstruction of ToothSegNet is the most complete, others have unreasonable disconnected fragments. In the frontal view, ToothSegNet results in the most smooth and natural reconstruction at the tooth connection, whereas UCtransNet leads to unnatural connections, U-Net and U-Net++ both end up with strange textures illustrated in a yellow box and a red box respectively. In the left view, one can observe that the reconstruction of UCTransNet, U-Net, and U-Net++ all contains floating tooth fragments, violating the common sense.
%本文方法重建的3D
%重建的病例是较为典型的困难病例，在萌出牙的slices中图像表现得都较为模糊，在牙冠区域的slices中，对比度差异较大，有较显著的金属伪影。
%MESH更加精确，没有重建出额外的信息，从（a）可以看出，本文方法对于牙根的重建最为完整，UCtrasnet、U-net、Un-e++t方法重建的有不相连的细碎，U-Net重建的mesh有漂浮的不相连的牙齿碎片；从（b）可以看出，本文方法在牙齿连接处最为平滑且自然，UCTRANSnet重建的有不自然的连接，Unet黄色区域与Unet++红色区域牙齿上重建出奇怪的纹理；从（c）可以看出，UCtransNet和Unet和Unet++重建出的mesh都有漂浮在空中的牙齿碎片，且不符合常理。总体来说，本文方法重建的mesh最为自然，最接近CBCT牙齿的原始形态。从而可以证明本文方法分割的精确性。

\subsection{Ablation study} 
Table 2 records the ablation study results for the degradation simulation module, structural constraint loss, and the CCF module. One can observe that all three designs are effective for segmentation, calling for the extension of this degradation simulation approach to other computer vision tasks. Also, we compare the performance of the structural constraint loss with the traditional SSIM loss \cite{qin2019basnet} as shown in Fig. 3, demonstrating that our constraint loss leads to preciser segmentation.

\begin{table}[t]
\centering
\caption{The objective metrics for ablation studies.}
\label{fig4}
\begin{tabular}{ccc|cccc} 
\hline
DS      & SC      & CCF~  & ~   IoU       &     Dice        & Recall         & Precision~                                                                                                                                                                                                                         \\ 
\hline\hline
$\surd$ & $\surd$ & $\surd$ & \ \ \textbf{90.21} & \textbf{94.85} & 97.77          & \textbf{92.11}                                                                                                                                                                                                                     \\
        & $\surd$ & $\surd$ & \ \ 89.90          &   94.68        & 97.79 & 91.76                                                                                                                                                                                                                              \\
$\surd$ &         & $\surd$ &\ \  89.67          &   94.55        & \textbf{97.93}          & 91.40                                                                                                                                                                                                                              \\
$\surd$ & $\surd$    &     & \ \  88.55       &       93.03         &    97.31            &       90.77                                                                                                                                                                                                                                        \\ 
\hline
\multicolumn{7}{l}{\begin{tabular}[c]{@{}l@{}}"DS" denotes~the degradation simulation module;\\"SC" denotes~the structural constraint loss;\\"CCF" denoetes~the channel-wise cross fusion module.~\end{tabular}}    \\
\hline

\end{tabular}
\vspace{-15pt} %调整图片与上文的垂直距离
\end{table}
\section{Conclusion}
\label{sec:conclusion}
Accurate tooth CBCT image segmentation is essential for the clinics in orthodontics. In this work, we combine the strengths of the designed degradation simulation module, CCF, and the structure constraint loss to provide a precise and robust automatic CBCT tooth segmentation. With in-depth analysis and ablation study, we show that the proposed ToothSegNet obtains better segmentation results on tooth datasets than the other six advanced medical image segmentation methods in terms of quantitative evaluation and visualization. In the future, we will deploy our work into clinical usage.

\bibliographystyle{IEEEbib}
\bibliography{strings,ref}

\end{document}